\documentclass{cimento}
\usepackage[centertags]{amsmath}
\allowdisplaybreaks[1]
\usepackage{amsbsy}
\usepackage{amsfonts}
\usepackage{amssymb}
\usepackage[dvips]{graphicx}
\usepackage{cite}

\title{The GSI Time Anomaly: Facts and Fiction}

\author{C. Giunti\from{ins:x}}

\instlist{\inst{ins:x} INFN, Sezione di Torino, Via P. Giuria 1, I--10125 Torino, Italy}

\begin{document}

\maketitle

\begin{abstract}
The claims
that the GSI time anomaly is due to the mixing of neutrinos
in the final state of the observed electron-capture decays of hydrogen-like heavy ions
are refuted
with the help of an analogy with
a double-slit experiment.
It is a consequence of causality.
It is shown that
the GSI time anomaly may be caused by
quantum beats due to the existence of two
coherent energy levels of the decaying ion
with an extremely small energy splitting
(about $6\times10^{-16}\,\text{eV}$)
and relative probabilities having a ratio of about 1/99.
\\[0.5cm]
\centerline{\textsf{
La Thuile 2009, Les Rencontres de Physique de La Vallee d'Aoste
}}
\\
\centerline{\textsf{
1-7 March 2009, La Thuile, Aosta Valley, Italy
}}
\null
\vspace{-0.9cm}
\null
\end{abstract}

A GSI experiment
\cite{0801.2079}
observed an anomalous oscillatory time modulation
of the electron-capture decays
\begin{align}
\null & \null
{}^{140}\text{Pr}^{58+} \to {}^{140}\text{Ce}^{58+} + \nu_{e}
\,,
\label{01a}
\\
\null & \null
{}^{142}\text{Pm}^{60+} \to ^{142}\text{Nd}^{60+} + \nu_{e}
\,.
\label{01b}
\end{align}
The hydrogen-like ions ${}^{140}\text{Pr}^{58+}$ and ${}^{142}\text{Pm}^{60+}$
were produced by fragmentation of a beam of ${}^{152}\text{Sm}$ with 500-600 MeV energy per nucleon on a ${}^{9}\text{Be}$ target
and stored in the ESR cooler-storage ring where they circulated with a frequency of about 2 MHz
and were monitored by Schottky Mass Spectrometry.
The electron capture data are fitted by an oscillatory decay rate with a period
$ T \simeq 7 \, \text{s} $ and an amplitude $ A \simeq 0.2 $ \cite{0801.2079}.

It has been proposed
\cite{0801.2079,0805.0435,0801.2121,0801.3262}
that the GSI anomaly is due to the interference of the massive neutrinos
which compose the final electron neutrino state,
\begin{equation}
| \nu_{e} \rangle = \cos\!\vartheta | \nu_{1} \rangle + \sin\!\vartheta | \nu_{2} \rangle
\,,
\label{nue}
\end{equation}
where $\vartheta$ is the solar mixing angle
(see Refs.~\cite{hep-ph/9812360,hep-ph/0202058,hep-ph/0405172,hep-ph/0606054,Giunti-Kim-2007,0704.1800}).

In order to assess the viability of this explanation of the GSI anomaly,
it is necessary to understand that interference is the result of the superposition of two or more waves
\cite{0805.0431}.
If the waves come from the same source,
interference can occur if the waves evolve different phases by propagating through different paths.
Therefore, interference occurs after wave propagation,
not at the wave source.
In the case of the GSI experiment,
there cannot be any interference effect of
$\nu_{1}$ and $\nu_{2}$ in the electron-capture decays (\ref{01a}) and (\ref{01b}),
which are the sources of $\nu_{1}$ and $\nu_{2}$.

\begin{figure}[t!]
\begin{center}
\includegraphics[clip, bb=68 551 549 757, width=\linewidth]{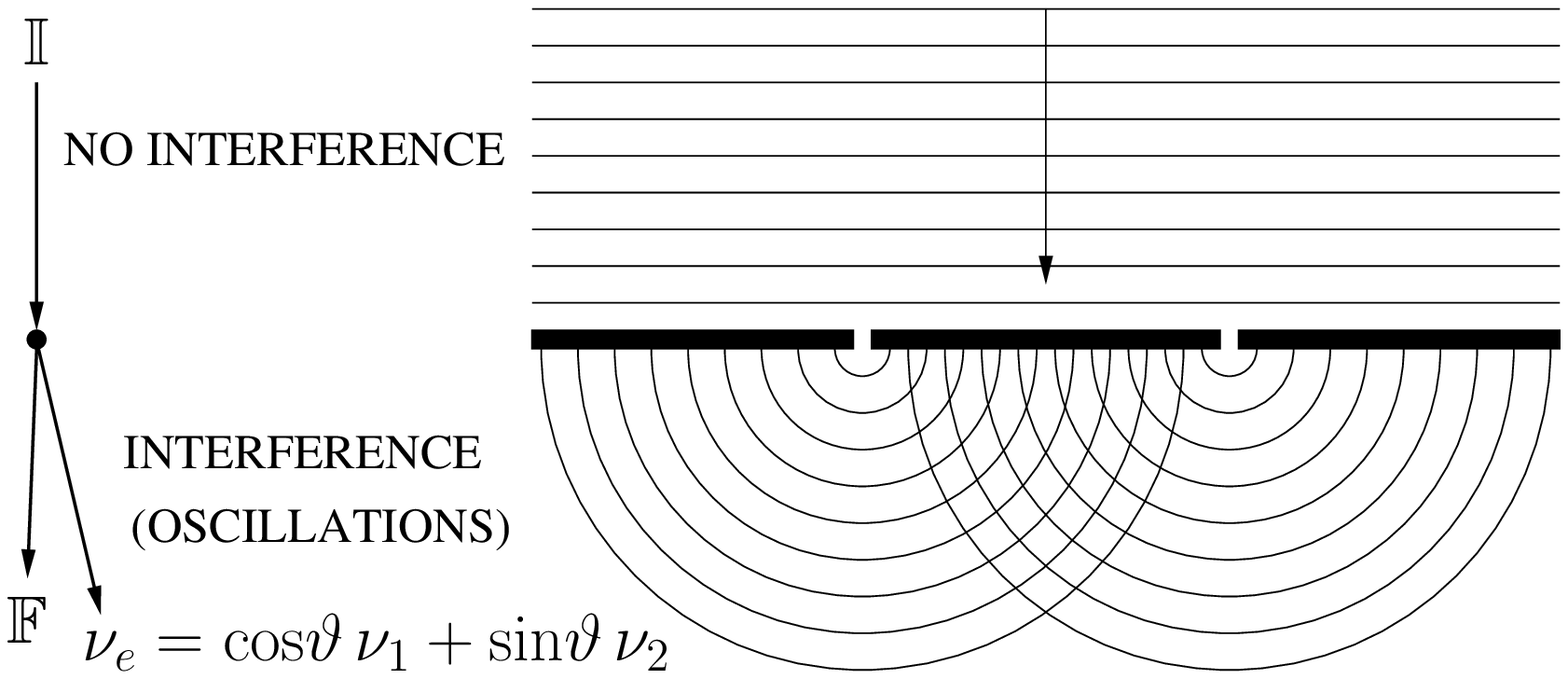}
\end{center}
\null\vspace{-1.0cm}\null
\caption{ \label{slit-1}
Analogy between the electron-capture decay process (\ref{003}) and a double-slit interference experiment.
}
\end{figure}

Let us illustrate these remarks through an analogy with
the well-known double-slit interference experiment with classical or quantum waves
depicted in Fig.~\ref{slit-1}.
In a double slit experiment an
incoming plane wave packet hits a barrier with two tiny holes,
generating two outgoing spherical wave packets which propagate on the other side of the barrier.
The two outgoing waves are coherent,
since they are created with the same initial phase in the two holes.
Hence, the intensity after the barrier,
which is proportional to the squared modulus of the sum of the two outgoing waves,
exhibits interference effects.
The interference depends on the different path lengths of the two outgoing spherical waves
after the barrier.

For the analogy with the double-slit experiment,
let us write schematically an electron-capture decay process of the type in Eqs.~(\ref{01a}) and (\ref{01b}) as
\begin{equation}
\mathbb{I} \to \mathbb{F} + \nu_{e}
\,.
\label{003}
\end{equation}
Taking into account the neutrino mixing in Eq.~(\ref{nue}),
we have two different decay channels:
\begin{equation}
\mathbb{I} \to \mathbb{F} + \nu_{1}
\,,
\qquad
\mathbb{I} \to \mathbb{F} + \nu_{2}
\,.
\label{004}
\end{equation}
The initial state in the two decay channels
is the same.
In our analogy with the double-slit experiment,
the initial state $\mathbb{I}$ is analogous to
the incoming wave packet.
The two final states
$ \mathbb{F} + \nu_{1} $
and
$ \mathbb{F} + \nu_{2} $
are analogous to the two outgoing wave packets.
Different weights of $ \nu_{1} $ and $ \nu_2 $ production
due to $ \vartheta \neq \pi/4 $
correspond to different sizes of the two holes in the barrier.

In the analogy,
the decay rate of $\mathbb{I}$ corresponds to the fraction of intensity of the incoming wave
which crosses the barrier,
which depends only on the sizes of the holes.
It does not depend on the interference effect which occurs after the wave has passed through the barrier.
In a similar way,
the decay rate of $\mathbb{I}$ cannot depend on the interference of $\nu_{1}$ and $\nu_{2}$
which occurs after the decay has happened.

Of course,
flavor neutrino oscillations caused by the interference of $\nu_{1}$ and $\nu_{2}$
can occur after the decay of $\mathbb{I}$,
in analogy with the occurrence of interference of the outgoing waves in the double-slit experiment,
regardless of the fact that the decay rate is
the incoherent sum of the rates of production of $\nu_{1}$ and $\nu_{2}$
and the fraction of intensity of the incoming wave
which crosses the barrier is the incoherent sum of the fractions of intensity of the incoming wave which
pass trough the two holes.

The above argument
is a simple consequence of causality:
the interference of $\nu_{1}$ and $\nu_{2}$ occurring after the decay
cannot affect the decay rate.

Causality is explicitly violated in Ref.~\cite{0805.0435},
where the decaying ion is described by a wave packet,
but it is claimed that there is a selection of the momenta of the ion
caused by a final neutrino momentum splitting due to the mass difference
of $\nu_{1}$ and $\nu_{2}$.
This selection violates causality.
In the double-slit analogy,
the properties of the outgoing wave packets are determined by the properties
of the incoming wave packet,
not vice versa.
In a correct treatment, all the
momentum distribution of the wave packet of the ion contributes to the decay,
generating appropriate neutrino wave packets.

The authors of Refs.\cite{0801.2121,0801.3262}
use a different approach:
they calculate the decay rate with the final neutrino state
\begin{equation}
| \nu \rangle
=
\sum_{k=1}^{3} | \nu_{k} \rangle
\,.
\label{005}
\end{equation}
This state is different from the standard electron neutrino state,
which is given by
\begin{equation}
| \nu_{e} \rangle
=
\sum_{k=1}^{3} U_{ek}^* \, | \nu_{k} \rangle
\,,
\label{021}
\end{equation}
where $U$ is the mixing matrix
(in the two-neutrino mixing approximation of Eq.~(\ref{nue}),
$U_{e1}=\cos\vartheta$,
$U_{e2}=\sin\vartheta$, and
$U_{e3}=0$).
It is not even properly normalized
to describe one particle ($\langle\nu|\nu\rangle=3$).
Moreover,
it leads to a decay rate which
is different from the standard decay rate,
given by the incoherent sum of the rates of decay into
the different massive neutrinos final states weighted by the
corresponding element of the mixing matrix
\cite{Shrock:1980vy,McKellar:1980cn,Kobzarev:1980nk,Shrock:1981ct,Shrock:1981wq}.
The decay rate is given by the integral over the phase space
of the decay probability
\begin{equation}
P_{\mathbb{I} \to \mathbb{F} + \nu}
=
| \langle \nu, \mathbb{F} | \mathsf{S} | \mathbb{I} \rangle |^2
=
\left| \sum_{k=1}^{3} \langle \nu_{k}, \mathbb{F} | \mathsf{S} | \mathbb{I} \rangle \right|^2
\,,
\label{011}
\end{equation}
where $\mathsf{S}$ is the S-matrix operator.
Considering the S-matrix operator at first order in perturbation theory,
\begin{equation}
\mathsf{S}
=
1
- i \int \text{d}^{4}x \, \mathcal{H}_{W}(x)
\,,
\label{012}
\end{equation}
with the effective four-fermion interaction Hamiltonian
\begin{align}
\mathcal{H}_{W}(x)
= 
\null & \null
\frac{G_F}{\sqrt{2}} \, \cos\theta_{\text{C}}
\,
\overline{\nu_{e}}(x) \gamma_{\rho} ( 1 - \gamma^5 ) e(x)
\,
\overline{n}(x) \gamma^{\rho} ( 1 - g_A \gamma^5 ) p(x)
\nonumber
\\
=
\null & \null
\frac{G_F}{\sqrt{2}} \, \cos\theta_{\text{C}}
\,
\sum_{k=1}^{3} \! U_{ek}^{*} \overline{\nu_{k}}(x) \gamma_{\rho} ( 1 - \gamma^5 ) e(x)
\,
\overline{n}(x) \gamma^{\rho} ( 1 - g_A \gamma^5 ) p(x)
\,,
\label{013}
\end{align}
where $\theta_{\text{C}}$ is the Cabibbo angle,
one can write the matrix elements in Eq.~(\ref{011}) as
\begin{equation}
\langle \nu_{k}, \mathbb{F} | \mathsf{S} | \mathbb{I} \rangle
=
U_{ek}^{*} \mathcal{M}_{k}
\,,
\label{014}
\end{equation}
with
\begin{equation}
\mathcal{M}_{k}
=
\frac{G_F}{\sqrt{2}} \, \cos\theta_{\text{C}}
\,
\langle \nu_{k}, \mathbb{F} |
\overline{\nu_{k}}(x) \gamma_{\rho} ( 1 - \gamma^5 ) e(x)
\,
\overline{n}(x) \gamma^{\rho} ( 1 - g_A \gamma^5 ) p(x)
| \mathbb{I} \rangle
\,.
\label{015}
\end{equation}
Therefore,
the decay probability is given by
\begin{equation}
P_{\mathbb{I} \to \mathbb{F} + \nu}
=
\left| \sum_{k=1}^{3} U_{ek}^{*} \mathcal{M}_{k} \right|^2
\,.
\label{016a}
\end{equation}
This decay probability is different from the standard one
\cite{Shrock:1980vy,McKellar:1980cn,Kobzarev:1980nk,Shrock:1981ct,Shrock:1981wq},
which is obtained by summing incoherently over the probabilities of decay
into the different massive neutrinos final states weighted by the
corresponding element of the mixing matrix:
\begin{equation}
P
=
\sum_{k=1}^{3} |U_{ek}|^2 \left| \mathcal{M}_{k} \right|^2
\,.
\label{016b}
\end{equation}

The analogy with the double-slit experiment
and the causality argument discussed above support the correctness of the standard
decay probability $P$.
Another argument against the decay probability $P_{\mathbb{I} \to \mathbb{F} + \nu}$
is that in the limit of massless neutrinos it does not
reduce to the decay probability in the Standard Model,
\begin{equation}
P_{\text{SM}}
=
\left| \mathcal{M}_{\text{SM}} \right|^2
\,,
\label{016}
\end{equation}
with
\begin{equation}
\mathcal{M}_{\text{SM}}
=
\frac{G_F}{\sqrt{2}} \, \cos\theta_{\text{C}}
\,
\langle \mathbb{F}, \nu_{e}^{\text{SM}} |
\overline{\nu_{e}^{\text{SM}}}(x) \gamma_{\rho} ( 1 - \gamma^5 ) e(x)
\,
\overline{n}(x) \gamma^{\rho} ( 1 - g_A \gamma^5 ) p(x)
| \mathbb{I} \rangle
\,,
\label{017}
\end{equation}
where $\nu_{e}^{\text{SM}}$ is the SM massless electron neutrino.
Indeed,
for the matrix elements $\mathcal{M}_{k}$ we have
\begin{equation}
\mathcal{M}_{k}
\xrightarrow[m_{k}\to0]{}
\mathcal{M}_{\text{SM}}
\,,
\label{018}
\end{equation}
leading to
\begin{equation}
P_{\mathbb{I} \to \mathbb{F} + \nu}
\xrightarrow[m_{k}\to0]{}
\left| \mathcal{M}_{\text{SM}} \right|^2
\left| \sum_{k=1}^{3} U_{ek}^{*} \right|^2
\,.
\label{019}
\end{equation}
This is different from the SM decay probability in Eq.~(\ref{016}).
Notice that the contribution of the elements
of the mixing matrix should disappear automatically
in the limit $m_{k}\to0$.
In fact, even in the SM one can define the three massless flavors neutrinos
$\nu_{e}$,
$\nu_{\mu}$,
$\nu_{\tau}$
as arbitrary unitary linear combinations of three massless neutrinos
$\nu_{1}$,
$\nu_{2}$,
$\nu_{3}$.
However,
all physical quantities are independent of such an arbitrary transformation.

We conclude that the state in Eq.~(\ref{005})
does not describe the neutrino emitted in an electron-capture decay process of the type in Eq.~(\ref{003})
and Refs.\cite{0801.2121,0801.3262} are flawed.

The correct normalized state
($\langle\nu_{e}|\nu_{e}\rangle=1$)
which describes
the electron neutrino emitted in an electron-capture decay processes of the type in Eq.~(\ref{003})
is \cite{hep-ph/0608070,Giunti-Kim-2007}
\begin{align}
| \nu_{e} \rangle
=
\null & \null
\left( \sum_{j}  | \langle \nu_{j}, \mathbb{F} | \mathsf{S} | \mathbb{I} \rangle |^2 \right)^{-1/2}
\sum_{k=1}^{3} | \nu_{k} \rangle \, \langle \nu_{k}, \mathbb{F} | \mathsf{S} | \mathbb{I} \rangle
\nonumber
\\
=
\null & \null
\left( \sum_{j}  | U_{ej} |^2 | \mathcal{M}_{j} |^2 \right)^{-1/2}
\sum_{k=1}^{3} U_{ek}^{*} \mathcal{M}_{k} \, | \nu_{k} \rangle
\,.
\label{121}
\end{align}
In experiments which are not sensitive to the differences of the neutrino masses,
as neutrino oscillation experiments,
we can approximate
$
\mathcal{M}_{k} \simeq \overline{\mathcal{M}}
$
and the state (\ref{121})
reduces to the standard electron neutrino state in Eq.~(\ref{021})
(apart for an irrelevant phase $\overline{\mathcal{M}}/|\overline{\mathcal{M}}|$).

With the electron neutrino state in Eq.~(\ref{121}),
the decay probability is given by
\begin{equation}
P_{\mathbb{I} \to \mathbb{F} + \nu_{e}}
=
| \langle \nu_e, \mathbb{F} | \mathsf{S} | \mathbb{I} \rangle |^2
=
\sum_{k=1}^{3} \left| \langle \nu_{k}, \mathbb{F} | \mathsf{S} | \mathbb{I} \rangle \right|^2
=
\sum_{k=1}^{3} |U_{ek}|^2 \left| \mathcal{M}_{k} \right|^2
\,.
\label{123}
\end{equation}
This is the correct standard result in Eq.~(\ref{016b}):
the decay probability is given by the incoherent sum over the probabilities of decay
into different massive neutrinos weighted by the
corresponding element of the mixing matrix.

Using Eq.~(\ref{018}) and the unitarity of the mixing matrix,
one can also easily check that $P_{\mathbb{I} \to \mathbb{F} + \nu_{e}}$
reduces to $P_{\text{SM}}$ in Eq.~(\ref{016})
in the massless neutrino limit.

\begin{figure}[t!]
\begin{center}
\includegraphics[clip, bb=69 477 553 756, width=\linewidth]{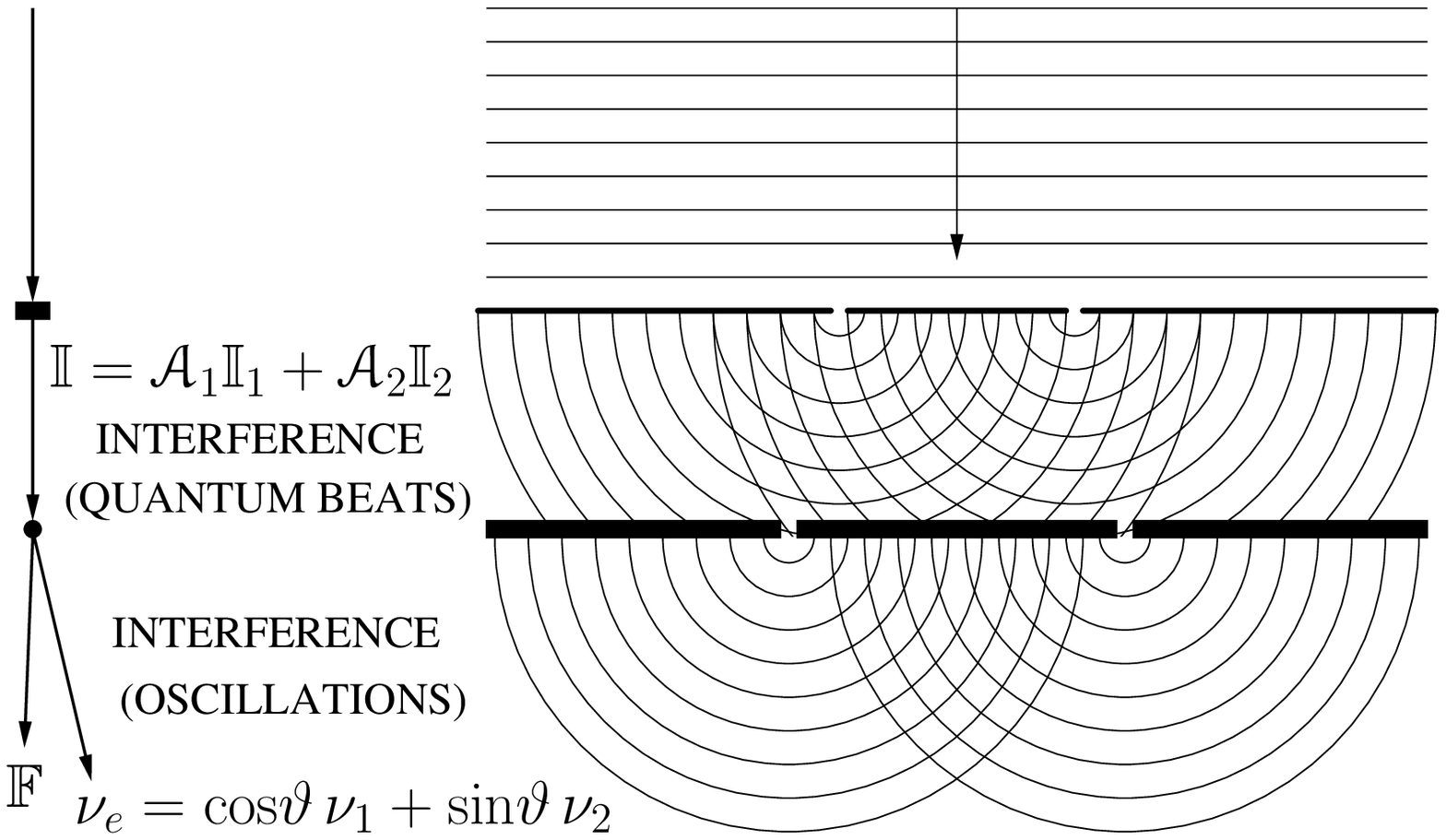}
\end{center}
\null\vspace{-1.0cm}\null
\caption{ \label{slit-2}
Analogy between quantum beats in the electron-capture decay process (\ref{003}) and a double-slit interference experiment with two coherent sources.
}
\end{figure}

Although the GSI time anomaly cannot be due to effects of neutrino mixing
in the final state of the electron-capture process,
it can be due to interference effects in the initial state.
In fact,
there could be an interference between two coherent energy states of the
decaying ion which produces quantum beats
(see, for example, Ref.~\cite{Carter-Huber-2000}).
Also in this case we can draw an analogy with a double-slit experiment.
However, we must change the setup,
considering the double-slit experiment with two coherent sources of waves
depicted in Fig.~\ref{slit-2}.
The two coherent sources are produced by an incoming plane wave packet hitting a first barrier with two tiny holes.
The two coherent outgoing waves interfere in the space between the first and the second barrier.
The interference at the holes in the second barrier
induces a modulation of the intensity which crosses the barrier.
The role of causality is clear:
the interference effect is due to the different phases
of the two coherent incoming waves at the holes in the second barrier,
which have developed during the propagation of the two waves
along different path lengths between the two barriers.
Analogously,
quantum beats in the GSI experiment can be due to
interference of two coherent energy states of the
decaying ion which develop different phases before the decay.
The two coherent energy states could be produced in the creation process of the ion,
which in GSI occurs through fragmentation of a beam of heavier ions on a target \cite{0801.2079},
as illustrated in Fig.~\ref{slit-2}.
Alternatively,
the two coherent energy states could be due to interactions of the decaying ion in the storage ring.

The quantum mechanical description of quantum beats is rather simple.
If the two energy states of the decaying ion
$\mathbb{I}_{1}$ and $\mathbb{I}_{2}$
are produced at the time $t=0$ with amplitudes
$\mathcal{A}_{1}$ and $\mathcal{A}_{2}$
(with $ |\mathcal{A}_{1}|^2 + |\mathcal{A}_{2}|^2 = 1 $),
we have
\begin{equation}
| \mathbb{I}(t=0) \rangle
=
\mathcal{A}_{1} \, | \mathbb{I}_{1} \rangle
+
\mathcal{A}_{2} \, | \mathbb{I}_{2} \rangle
\,.
\label{201}
\end{equation}
Assuming, for simplicity, that the two states
with energies
$E_{1}$ and $E_{2}$
have the same decay rate $\Gamma$,
at the time $t$
we have
\begin{equation}
| \mathbb{I}(t) \rangle
=
\left(
\mathcal{A}_{1} \, e^{ - i E_{1} t } \, | \mathbb{I}_{1} \rangle
+
\mathcal{A}_{2} \, e^{ - i E_{2} t } \, | \mathbb{I}_{2} \rangle
\right)
e^{ - \Gamma t / 2 }
\,.
\label{202}
\end{equation}
The probability of electron capture at the time $t$ is given by
\begin{align}
P_{\text{EC}}(t)
=
\null & \null
| \langle \nu_{e}, \mathbb{F} | \mathsf{S} | \mathbb{I}(t) \rangle |^2
\nonumber
\\
=
\null & \null
\left[
1
+
A \, \cos\!\left( \Delta{E} t + \varphi \right)
\right]
\overline{P}_{\text{EC}}
\,
e^{ - \Gamma t }
\,.
\label{203}
\end{align}
where
where $\mathsf{S}$ is the S-matrix operator,
$ A \equiv 2 |\mathcal{A}_{1}| |\mathcal{A}_{2}| $,
$ \Delta{E} \equiv E_{2} - E_{1} $,
\begin{equation}
\overline{P}_{\text{EC}}
=
| \langle \nu_{e}, \mathbb{F} | \mathsf{S} | \mathbb{I}_{1} \rangle |^2
=
| \langle \nu_{e}, \mathbb{F} | \mathsf{S} | \mathbb{I}_{2} \rangle |^2
\,,
\label{204}
\end{equation}
and $\varphi$ is a constant phase which takes into account possible phase differences of
$\mathcal{A}_{1}$ and $\mathcal{A}_{2}$
and of
$ \langle \nu_{e}, \mathbb{F} | \mathsf{S} | \mathbb{I}_{1} \rangle $
and
$ \langle \nu_{e}, \mathbb{F} | \mathsf{S} | \mathbb{I}_{2} \rangle $.

The fit of GSI data presented in Ref.~\cite{0801.2079} gave
\begin{align}
\null & \null
\Delta{E}({}^{140}\text{Pr}^{58+})
=
( 5.86 \pm 0.07 ) \times 10^{-16} \, \text{eV}
\,,
\null & \null
\quad
\null & \null
A({}^{140}\text{Pr}^{58+})
=
0.18 \pm 0.03
\,,
\label{211}
\\
\null & \null
\Delta{E}({}^{142}\text{Pm}^{60+})
=
( 5.82 \pm 0.18 ) \times 10^{-16} \, \text{eV}
\,,
\null & \null
\quad
\null & \null
A({}^{142}\text{Pm}^{60+})
=
0.23 \pm 0.04
\,.
\label{212}
\end{align}
Therefore,
the energy splitting is extremely small
and the oscillation amplitude $A$ is significantly smaller than one.

The authors of Ref.~\cite{0801.2079}
noted that the splitting of the two hyperfine $1s$ energy levels
of the electron is many order of magnitude too large
(and the contribution to the decay of one of the two states is
suppressed by angular momentum conservation).
It is difficult to find a mechanism which produces a smaller
energy splitting.
Furthermore,
since the amplitude $A\simeq0.2$ of the interference is rather small,
it is necessary to find a mechanism which
generates coherently the states
$\mathbb{I}_{1}$ and $\mathbb{I}_{2}$
with probabilities
$|\mathcal{A}_{1}|^2$ and $|\mathcal{A}_{2}|^2$
having a ratio of about 1/99!

An important question is if the coherence of $\mathbb{I}_{1}$ and $\mathbb{I}_{2}$ is preserved
during the decay time.
Since the measuring apparatus monitors the ions through elastic electromagnetic interactions
with a frequency of
the order of the revolution frequency in the ESR storage ring, about 2 MHz,
the coherence can be preserved only if the interaction with the measuring apparatus does not
distinguish between the two states.
In this case the interaction is coherent,
i.e. the two states suffer the same phase shift.
Since the energy splitting $ \Delta{E} $ is extremely small,
I think that coherence is maintained for a long time if $\mathbb{I}_{1}$ and $\mathbb{I}_{2}$
have the same electromagnetic properties.

In conclusion, I have shown that
the GSI time anomaly cannot be due to neutrino mixing in the final state
of the observed electron-capture decays of the hydrogen-like ${}^{140}\text{Pr}^{58+}$ and ${}^{142}\text{Pm}^{60+}$
ions.
The argument has been clarified through an analogy with
a double-slit experiment, emphasizing that it is a consequence of causality \cite{0805.0431}.
I have explained the reasons why the claim in Refs.~\cite{0805.0435,0801.2121,0801.3262}
that the GSI time anomaly is due to the mixing of neutrinos
in the final state of the electron-capture process
is incorrect (see also Refs.~\cite{0804.1099,0808.2389}).
I have also shown that
the GSI time anomaly may be due to
quantum beats due to the existence of two
coherent energy levels of the decaying ion.
However,
since the required energy splitting is extremely small
(about $6\times10^{-16}\,\text{eV}$)
and the two energy levels must be produced with relative probabilities having a ratio of about $1/99$,
it is very difficult to find an appropriate mechanism.

\end{document}